\documentclass{icsm}         
\usepackage{graphicx}
\usepackage{bm}
\usepackage{amsmath}
\usepackage{amsfonts}
\usepackage{amssymb}
\usepackage{amsthm}

\usepackage[dvips]{epsfig}

\begin{document}
\title{Deformed Bosons: Combinatorics of Normal Ordering}
\authori{P. Blasiak$^{a,b}$, A. Horzela$^{b}$,}
\addressi{$^{a}$Laboratoire de Physique Th\'eorique des Liquides,
Universit\'e Pierre et Marie Curie,\\
Tour 24 - 2e \'et., 4 Pl.Jussieu, F 75252 Paris Cedex 05, France
\\
$^{b}$H.Niewodnicza\'nski Institute of Nuclear Physics, Polish Academy of Sciences,\\
ul. Eliasza-Radzikowskiego 152, PL 31342 Krakow, Poland\\
e-mail: blasiak@lptl.jussieu.fr, andrzej.horzela@ifj.edu.pl}
\authorii{K. A. Penson$^{a}$ and A. I. Solomon$^{a,c}$}
\addressii{$^{a}$Laboratoire de Physique Th\'eorique des Liquides,
Universit\'e Pierre et Marie Curie,\\
Tour 24 - 2e \'et., 4 Pl.Jussieu, F 75252 Paris Cedex 05, France\\
$^{c}$The Open University, Physics and Astronomy Department,
\\Milton Keynes MK7 6AA, United Kingdom
\\email: penson@lptl.jussieu.fr,  a.i.solomon@open.ac.uk}
\authoriii{}      \addressiii{}
%
\headauthor{P. Bl{}asiak et al.}
\headtitle{Deformed Bosons: Combinatorics of Normal Ordering}
\lastevenhead{ P. Bl{}asiak et al.: Deformed Bosons: Combinatorics
of Normal Ordering}
%
\maketitle
\begin{abstract}
We solve the normal ordering problem for $(A^{\dag}A)^n$ where $A$
and $A^\dag$ are one mode deformed ($[A,A^\dag]=[N+1]-[N]$)
bosonic ladder operators. The solution generalizes results known
for canonical bosons. It involves combinatorial polynomials in the
number operator $N$ for which the generating function and explicit
expressions are found. Simple deformations provide examples of the
method.
\end{abstract}
\section{Introduction}
We consider the general deformation of the boson algebra
\cite{Solomon} in the form
\begin{equation}
\label{algebra}
\begin{array}{lcccr}
[A,N]=A,&~~~&[A^\dag,N]=-A^\dag,&~~~&[A,A^\dag]=[N+1]-[N].
\end{array}
\end{equation}
In the above $A$ and $A^\dag$ are annihilation and creation
operators, respectively, while the number operator $N$ counts
particles. It is defined in the Fock basis as
$N|n\rangle=n|n\rangle$ and commutes with $A^\dag A$. Because of
that in any representation of (\ref{algebra}) $N$ can be written
in the form $A^\dag A = [N]$, where $[N]$ denotes an arbitrary
function of $N$, usually called the ''box'' function. For general
considerations we do not assume any  realization of the number
operator $N$ and we treat it as an independent element of the
algebra.  Moreover, we do not assume any particular form of the
''box'' function $[N]$. Special cases, like $so(3)$ or $so(2,1)$
algebras, will provide  examples that show how such a general
approach simplifies if an algebra and its realization are chosen.

In this note we give the solution to the problem of the normal
ordering of a monomial $(A^\dag A)^n$ in  deformed annihilation
and creation operators. A classical result due to J. Katriel
\cite{Katriel} is that for canonical bosons $a$ and $a^\dag$, {\it
i.e.,} for the Heisenberg-Weyl algebra,  we have
\begin{eqnarray}
\label{S} (a^\dag a)^n=\sum_{k=1}^n \mathbb{S}_{n,k} (a^\dag)^k
a^k,\ \ \ \ \ \ \ \ {\rm{(for\ canonical\ bosons)}}
\end{eqnarray}
where $\mathbb{S}_{n,k}$ are the Stirling numbers of the second
kind which count the number of partitions of an $n$ element set
into $k$ subsets. Namely, (\ref{S}) links the problem to
combinatorics. In the case of deformed bosons we cannot express
the monomial $(A^\dag A)^n$ as the combination of normally ordered
expressions in $A^\dag$ and $A$ only. This was found for
$q$-deformed bosons some time ago, \cite{KatrielKibler}, and
recently for the R-deformed Heisenberg algebra related to the
Calogero model, \cite{Burdik}. The number operator $N$ occurs in
the final formulae because moving creation operators to the left
cannot get rid of $N$ if it is assumed to be an independent
element of the algebra. It appears that in general we can look for
a solution of the form
\begin{eqnarray}
\label{P} (A^\dag A)^n=\sum_{k=1}^n \mathbb{P}_{n,k}(N)\
(A^\dag)^k A^k,
\end{eqnarray}
where the coefficients $\mathbb{P}_{n,k}(N)$ are functions of the
number operator $N$ which depend on the box function $[N]$.
Following \cite{KatrielKibler} we will call them {\em
operator-valued deformed Stirling numbers} or just {\em deformed
Stirling numbers}. In the sequel we give a comprehensive analysis
of this generalization. We give recurrences satisfied by
$\mathbb{P}_{n,k}(N)$ of (\ref{P}) and  construct their generating
functions. This will enable us to give $\mathbb{P}_{n,k}(N)$
explicitly and to demonstrate how the method works on examples of
simple Lie-type deformations of the canonical case.
\section{Recurrence relations}
One may check by induction that for each $k\geq 1$ the following
relation holds
\begin{eqnarray}\label{a}
[A^k,A^\dag]=([N+k]-[N])A^{k-1}.
\end{eqnarray}
Using this relation it is also easy to check by induction that deformed
Stirling numbers satisfy the recurrences\footnote[1]{The
recurrence relation Eq.(\ref{recurrence}) holds for all $n$ and
$k$ if one imposes the following ''boundary conditions'':
$\mathbb{P}_{i,j}(N)=0$ for $i=0$ or $j=0$ or $i<j$, except
$\mathbb{P}_{0,0}(N)=1$.}
\begin{eqnarray}
\label{recurrence}
\mathbb{P}_{n+1,k}(N)=\mathbb{P}_{n,k-1}(N)+([N]-[N-k])\mathbb{P}_{n,k}(N)\
\ \ \ \ \textstyle{{\rm for}\ 1<k<n}
\end{eqnarray}
with initial values
\begin{eqnarray}
\label{initial} \mathbb{P}_{n,1}(N)=([N]-[N-1])^{n-1}, \ \ \ \ \ \
\ \mathbb{P}_{n,n}(N)=1.
\end{eqnarray}
The proof can be deduced from the equalities
\begin{equation}
\label{polynomials}
\begin{array}{l}
\sum_{k=1}^{n+1}\mathbb{P}_{n+1,k}(N)\ (A^\dag)^k A^k=(A^\dag
A)^{n+1}=(A^\dag A)^n \ A^\dag A\nonumber
\\\\
=\sum_{k=1}^n \mathbb{P}_{n,k}(N)\ (A^\dag)^k\ A^k A^\dag\
A\stackrel{(\ref{a})}{=} \sum_{k=1}^n \mathbb{P}_{n,k}(N)\
(A^\dag)^k\left(A^\dag A+[N+k]-[N]\right)A^k
\\\\
\stackrel{(\ref{algebra})}{=}\sum_{k=2}^{n+1}\mathbb{P}_{n,k-1}(N)\
(A^\dag)^k A^k+\sum_{k=1}^n([N]-[N-k])\mathbb{P}_{n,k}(N)\
(A^\dag)^k A^k.
\end{array}
\end{equation}
It can be shown that each $\mathbb{P}_{n,k}(N)$ is a homogenous
polynomial of order $n-k$ in the variables $[N],...,[N-k]$. For a
polynomial box function one gets $\mathbb{P}_{n,k}(N)$ as a
polynomial in $N$.  As we shall show, a simple example of the
latter is the deformation of the  $so(3)$ or $so(2,1)$ Lie
algebras.

Note that for the ''box'' function $[N]=N+{\rm const}$ ({\it
i.e.}, for the canonical algebra) we get the ''ordinary'' Stirling
numbers \cite{Comtet} of the second kind which fulfill the
recurrence relation $\mathbb{S}_{n+1,k}=\mathbb{S}_{n,k-1}+k\
\mathbb{S}_{n,k}$ with initial values
$\mathbb{S}_{n,n}=\mathbb{S}_{n,1}=1$.
\section{Generating functions and general expressions}
We define the set of ordinary generating functions (OGF) of
polynomials $\mathbb{P}_{n,k}(N)$, for $k\geq 1$, in the form
\begin{eqnarray}\label{Pk}
P_{k}(x,N):=\sum_{n=k}^\infty \mathbb{P}_{n,k}(N)\ x^n.
\end{eqnarray}
The initial conditions (\ref{initial}) for $k=1$ give
\begin{eqnarray}\label{A}
P_1(x,N)=\frac{x}{1-([N]-[N-1])\ x}.
\end{eqnarray}
Using the recurrences (\ref{recurrence}) supplemented by
(\ref{initial}) one finds the relation
\begin{eqnarray}\label{B}
P_k(x,N)=\frac{x}{1-([N]-[N-k])\ x}\ P_{k-1}(x,N),\ \ \ \ \ {\rm
for}\ k>1,
\end{eqnarray}
for which the proof is provided by the equalities
\begin{equation}
\label{polynomials2}
\begin{array}{l}    P_k(x,N)\stackrel{(\ref{recurrence})}{=}\sum_{n=k}^\infty\left(\mathbb{P}_{n-1,k-1}(N)+([N]-[N-k])\ \mathbb{P}_{n-1,k}(N)\right)\ x^n
\\\\
=x\sum_{n=k}^\infty \mathbb{P}_{n-1,k-1}(N)\ x^{n-1}+([N]-[N-k])\
x\sum_{n=k}^\infty \mathbb{P}_{n-1,k}(N)\ x^{n-1}\nonumber
\\\\
=x\ P_{k-1}(x,N)+([N]-[N-k])\ x\ P_k (x,N).
\end{array}
\end{equation}
The expressions (\ref{A}) and (\ref{B}) give explicit formulae for
the OGF
\begin{eqnarray}\label{OGF}
P_k(x,N)=\prod_{j=1}^k \frac{x}{1-([N]-[N-j])\ x}.
\end{eqnarray}
For the canonical algebra this becomes in the OGF for Stirling
numbers:
\begin{eqnarray}
P_k(x)=\frac{x}{(1-x)(1-2x)\cdot...\cdot(1-kx)}.
\end{eqnarray}

The explicit knowledge of the OGF (\ref{OGF}) enables us to find
the $\mathbb{P}_{n,k}(N)$ in a compact form. As a rational
function of $x$ it can be expressed as a sum of partial fractions
\begin{eqnarray}
P_k(x,N)=x^k\ \sum_{r=1}^k\frac{\alpha_r}{1-([N]-[N-r])\ x},
\end{eqnarray}
where
\begin{eqnarray}
\label{alphas} \alpha_r=\frac{1}{\prod_{j=1,j\neq
r}^k\left(1-\frac{[N]-[N-j]}{[N]-[N-r]}\right)}.
\end{eqnarray}
From the definition (\ref{Pk}) we have $\mathbb{P}_{n,k}(N)=[x^n]\
P_k(x,N)$\footnote{This notation comes from combinatorics and
denotes the coefficient multiplying $x^n/n!$ in the formal Taylor
expansion of the generating function.}. Expanding the fractions in
the above equations and collecting the terms we get
$\mathbb{P}_{n,k}(N)=\sum_{r=1}^k\alpha_r([N]-[N-r])^{n-k}$.
Finally,  using (\ref{alphas}), we arrive at
\begin{eqnarray}
\mathbb{P}_{n,k}(N)=\sum_{r=1}^k\frac{([N]-[N-r])^{n-1}}{\prod_{j=1,j\neq
r}^k([N-j]-[N-r])}
\end{eqnarray}
where monotonicity of $[N]$ has been assumed. For the canonical case this yields the classical Stirling numbers $\mathbb{S}_{n,k} =
\sum_{r=1}^k(-1)^{k-r}\frac{r^n}{r!\ (k-r)!}$,  \cite{Comtet}.

The first four families of general deformed Stirling polynomials
may be obtained from  (\ref{P}) with a little help from standard
computer algebra tools, and  are
\begin{eqnarray}\nonumber
    \begin{array}{l}
\mathbb{P}_{1,1}(N)=1,\\
\\
\mathbb{P}_{2,1}(N)=[N]-[N-1],\\
\mathbb{P}_{2,2}(N)=1,\\
\\
\mathbb{P}_{3,1}(N)=([N]-[N-1])^2,\\
\mathbb{P}_{3,2}(N)=2[N]-[N-1]-[N-2],\\
\mathbb{P}_{3,3}(N)=1,\\
\\
\mathbb{P}_{4,1}(N)=([N]-[N-1])^3,\\
\mathbb{P}_{4,2}(N)=3[N]^2-3[N][N-1]-3[N][N-2]+[N-1]^2+[N-2][N-1]+[N-2]^2,\\
\mathbb{P}_{4,3}(N)=3[N]-[N-1]-[N-2]-[N-3],\\
\mathbb{P}_{4,4}(N)=1.
  \end{array}
\end{eqnarray}

\section{Examples of simple deformations}
If we fix  the ''box'' function as $[N]=\mp\frac{N(N-1)}{2}$ then
(\ref{algebra}) gives the commutation relations of $so(3)$ and
$so(2,1)$ algebras, respectively. The first four families of
polynomials, satisfying $\mathbb{P}_{n,k}^{so(2,1)}(N)=(-1)^{n-k}\
\mathbb{P}_{n,k}^{so(3)}(N)$, because of (\ref{recurrence}) and
(\ref{initial}), are
\begin{eqnarray}\nonumber
    \begin{array}{lccl}
    \textbf{so(3)}&&&\textbf{so(2,1)}\\
    \\
    \mathbb{P}_{1,1}(N)=1&&&\mathbb{P}_{1,1}(N)=1\\
    \\
    \mathbb{P}_{2,1}(N)=-N+1&&&\mathbb{P}_{2,1}(N)=N-1\\
    \mathbb{P}_{2,2}(N)=1&&&\mathbb{P}_{2,2}(N)=1\\
    \\
    \mathbb{P}_{3,1}(N)=(N-1)^2&&&\mathbb{P}_{3,1}(N)=(N-1)^2\\
    \mathbb{P}_{3,2}(N)=-3N+4&&&\mathbb{P}_{3,2}(N)=3N-4\\
    \mathbb{P}_{3,3}(N)=1&&&\mathbb{P}_{3,3}(N)=1\\
    \\
    \mathbb{P}_{4,1}(N)=-(N-1)^3&&&\mathbb{P}_{4,1}(N)=(N-1)^3\\
    \mathbb{P}_{4,2}(N)=7N^2-19N+13&&&\mathbb{P}_{4,2}(N)=7N^2-19N+13\\
    \mathbb{P}_{4,3}(N)=-6N+10&&&\mathbb{P}_{4,3}(N)=6N-10\\
    \mathbb{P}_{4,4}(N)=1&&&\mathbb{P}_{4,4}(N)=1\\
    \end{array}
\end{eqnarray}

Up to now we have been considering $A$, $A^\dag$, and $N$ as
independent  elements of the algebra. Choosing a representation
and expressing $N$ in terms of $A$ and $A^\dag$ we arrive at
expressions like (\ref{P}) which can be normally ordered further.
We may consider a realization of  $so(2,1)$ in terms of
canonical operators $a$ and $a^\dag$
\begin{equation}
\label{so21}
\begin{array}{lcccr}
A = \frac{\displaystyle 1}{\displaystyle 2\sqrt{2}}aa &~~~& A^\dag
= \frac{\displaystyle 1}{\displaystyle 2\sqrt{2}}a^\dag a^\dag
&~~~& N = \frac{\displaystyle 1}{\displaystyle {4}}\left\{a^\dag,
a\right\}.
\end{array}
\end{equation}
\noindent The solution of the normal ordering problem for
$\left((a^\dag)^2a^2\right)^{n}$, \cite{Blasiak}, leads to
\begin{equation}
\label{genstirling1}
\begin{array}{lcr}
\left((a^\dag)^2a^2\right)^n &=& \sum\limits_{k=2}^{2n}
\mathbb{S}_{2,2}(n,k)a^{\dag k}a^k,
\end{array}
\end{equation}
\noindent where $\mathbb{S}_{2,2}(n,k)$ are generalized Stirling
numbers of the second kind defined by
\begin{equation}
\label{genstirling2}
\begin{array}{lcr}
\mathbb{S}_{2,2}(n,k)&=&\frac{\displaystyle{(-1)^k}}{\displaystyle
k!}\sum\limits_{p=2}^{k}(-1)^{p}
\left(\!\!\begin{array}{c}k\\p\end{array}\!\!\right)\left[p(p-1)\right]^n
\\
&=& \sum\limits_{l=0}^{n}
(-1)^{l}\left(\!\!\begin{array}{c}n\\l\end{array}\!\!\right)\mathbb{S}_{1,1}(2n-l,k).
\end{array}
\end{equation}
\noindent The numbers $\mathbb{S}_{2,2}(n,k)$ also have a
combinatorial interpretation which generalize partitions of a set
(classical Stirling numbers). Consider $2n$ distinguishable
objects anti-correlated in pairs. This can be realized by
colouring them in such a way that there are exactly two objects of
each colour, {\it i.e.} there are $n$ anti-correlated (coloured)
pairs. Now, numbers $\mathbb{S}_{2,2}(n,k)$ count the partitions
of that set into $k$ subsets with the restriction that
anti-correlated pairs are divided among different subsets, {\it
i.e.} all objects in a subset are of different colours. This means
that we restrict partitions of a $2n$ set by requiring
anti-correlation of pairs. This interpretation will be taken
further elsewhere \cite{Int}.

\noindent Terms in (\ref{genstirling1}) with $k$ even are directly
expressible by $A$ and $A^{\dag}$. If $k$ is odd then the factor
$a^{\dag}a$ may be moved to the left and with the convention
$\mathbb{S}_{2,2}(n,k)=0$ for $k>2n$ we get
\begin{equation}
\left((a^\dag)^2a^2\right)^n = \sum\limits_{k=1}^{n}
\left(\mathbb{S}_{2,2}(n,2k) + \mathbb{S}_{2,2}(n,2k+1)(a^\dag a -
2k)\right)a^{\dag 2k}a^{2k}.
\end{equation}
Finally, we get
\begin{equation}
(A^\dag A)^n=\sum\limits_{k=1}^{n}
8^{k-n}\left(\mathbb{S}_{2,2}(n,2k) + \mathbb{S}_{2,2}(n,2k+1)(2N
- 2k - 1/2)\right)A^{\dag k}A^{k}.
\end{equation}
which shows that the general solution involving higher order
polynomials in N can be simplified when a particular realization
of the algebra is postulated.

\section{Summary}

We considered the problem of the normal ordering of monomials
$(A^\dag A)^n$ for generally deformed boson creation and
annihilation operators.  The general solution involves some
combinatorial-like polynomials for which ordinary generating
functions and explicit formulas can be given. It was shown that
for particular realizations of the deformed algebra the solution
may be further simplified.

{\small P.B. wishes to thank the Polish Ministry of Scientific
Research and Information Technology for support under Grant no:
1P03B 051 26.}

\end{document}